# Ontological Matchmaking in Recommender Systems

Angela Bonifati
CNR, Italy
Angela.Bonifati@gmail.com

Giansalvatore Mecca
University of Basilicata, Italy
mecca@unibas.it

Domenica Sileo
University of Basilicata, Italy
Domenica.Sileo@gmail.com

Gianvito Summa
University of Basilicata, Italy
Gianvito.Summa@gmail.com

October 28, 2018

**Abstract**

The electronic marketplace offers great potential for the recommendation of supplies. In the so called recommender systems, it is crucial to apply matchmaking strategies that faithfully satisfy the predicates specified in the demand, and take into account as much as possible the user preferences. We focus on real-life ontology-driven matchmaking scenarios and identify a number of challenges, being inspired by such scenarios. A key challenge is that of presenting the results to the users in an understandable and clear-cut fashion in order to facilitate the analysis of the results. Indeed, such scenarios evoke the opportunity to rank and group the results according to specific criteria. A further challenge consists of presenting the results to the user in an asynchronous fashion, i.e. the 'push' mode, along with the 'pull' mode, in which the user explicitly issues a query, and displays the results. Moreover, an important issue to consider in real-life cases is the possibility of submitting a query to multiple providers, and collecting the various results. We have designed and implemented an ontology-based matchmaking system that suitably addresses the above challenges. We have conducted a comprehensive experimental study, in order to investigate the usability of the system, the performance and the effectiveness of the matchmaking strategies with real ontological datasets.

# 1  Introduction

The amount of information available on the Internet is enormously increasing and the need of fully leveraging such information to satisfy the users needs is becoming crucial. Recommender systems [1] have gained a lot of popularity as effective means to improve navigability of web sites and to help users and customers to quickly locate items of interest. Past works have been devoted to design effective matchmaking algorithms [9, 4], capable of yielding recommendations that best suit the users demands. Other recent recommendation tools, such as Yahoo! Vibes [12], aim at deploying a variety of recommendation models, applicable to a broad variety of domains and to large-scale networks, such as the Grid. Lately, [2, 8] have focused on the problem of making appropriate recommendation in social-tagging web sites and of designing recommendation workflows, respectively.

However, despite the rich literature of recommendation algorithms and tools, a few challenges still remain unaddressed. Such challenges arise in real-life Web 2.0 scenarios, in which more semantics is necessary to assist the user in the process of choosing the items that are likely to be of their interest. The limitations of previous solutions are indeed the fact that they are customized to a particular domain and/or hard-wired. In this paper, we present an ontology-driven recommender system, that is generic in nature and readily deployable in a Semantic Web environment. The key motivation behind our work is that recommender systems have to cope with real items distributed over the Web. The reasons for such belief are indeed numerous. First of all, it is more and more common to find ontology-based management systems (OntologyBMSs), that describe the properties of goods or services. OntologyBMSs are easy to set up by organizations that expose their items to the Internet, and may also be used internally by these organizations to facilitate search and storage. An OntologyBMS is a system that stores ontology schemas (TBox) and ontology instances (ABox) in a given data format. We have studied and implemented a recommendation platform, in which such OntologyBMSs are easily pluggable, by simply exporting a reference to their ontologies to an external registry. Another reason why ontologies are crucial in the matchmaking process, is due to the fact that they let exploit computer-usable definition of basic concepts in the domain and their relationships. As such, the ontologies are extremely useful in a distributed and heterogeneous scenarios in which different providers may offer the same goods with different (but equivalent) descriptions. In such scenarios, ontologies are essential to facilitate the processes of sharing, reusing and integrating information. OWL [14] has become the de facto standard for ontology development in diverse fields. Many OWL ontologies are available on the Web, identified by an URI, and there are also several well-known ontology libraries and ontology search engines (among which, SWOOGLE [16], Sindice [15] and Watson [17]).

However, applications are still built around a predetermined set of ontologies, that are well-understood [6]. We believe that a recommender system should be able to exploit the diversity of the ontology domains, by means of a flexible and generic architecture.

In the following, we conclude by summarizing the main contributions of our work.

## 1.1 Summary of contributions

We have designed and implemented an ontology-based matchmaking system, that leverages the semantics of OWL to yield recommendations to the users and satisfy their demands. The main contributions of our work can be summarized as follows:

- our system leverages the ontology axioms to build adequate matchmaking results and satisfy the user needs; moreover, the same semantic constructs are used to automatically deliver preferred results to the user by leveraging user profiles;

- we have designed and implemented a centralized architecture and a distributed architecture. The distributed version of our system exploits the Web services technology and the SOAP/XML protocol;

- the matchmaking algorithm that is the bulk of our system, inspired by [4], has been substantially extended and enhanced with knowledge elicitation. The extension takes into account the OWL constructs and turns to be extremely useful for result visualization in real-life Web 2.0 scenarios.

We have conducted an experimental study in which we investigate the effectiveness of the matchmaking and recommendation; the performance of the matchmaking algorithm with knowledge elicitation; and, finally, the efficiency of the distributed matchmaking, if compared with network latency time.

The paper is organized as follows. Section 2 discusses a motivating example, along with the centralized and the distributed architectures of our system, by highlighting the differences between the two. Section 3 discusses the matchmaking process, which constitutes the core of both architectures. Section 4 discusses the visualization and presentation enhancements on the results of the matchmaking process, and their usefulness. Section 5 presents an experimental study, showing both the effectiveness of the matchmaking processes, and the performance of the centralized and distributed architectures. Section 6 discusses the related approaches and techniques for recommender systems, and matchmaking systems. Finally, Section 7 concludes our work and presents future directions of investigation.

## 2 Overview of an Ontology-based Recommender System

In Section 2.1, we first present a motivating example, that shows the features and capabilities of our ontology-based recommender system at work. Then, in Section 2.2, we illustrate the characteristics of the architectures we have designed for our system.

### 2.1 A typical Purchase Use Case

Let us assume that a user is looking for a white-coloured laptop with at least 2-years warranty. While the user is posing the query through the browser, by filling a form, that presents other features, such as the cost and the operating system. However, the user may not be acquainted with the possible values of operating system or he may be not interested to a specific operating system. Similarly, he may not want to specify a constraint on the laptop cost at his first search. This may be due to the fact that the user is not yet clear on whether to buy a cheap or expensive laptop, until he can actually see the available laptop models.

These considerations brought us to design a matchmaking system, that is fully exploiting the semantics of ontologies to guide the user through the purchase process. In particular, our system can cover the main ontology axioms [6], including *equivalent classes*, *subclasses*, *object properties*, *datatype properties*, *functional properties* and *inverses*, as shown in the following OWL snippet capturing an ontology for the laptop purchase use case.

```
Class: Laptop
  EquivalentTo: {PortableComputer, MobileComputer}
DatatypeProperty: model
DatatypeProperty: warrantyYears
DatatypeProperty: colour
DatatypeProperty: cost
DatatypeProperty: operatingSystem
  SubClassOf: hasColour max 1
ObjectProperty: hasSerialNumber
  Inverses: isSerialNumberof
```

By inspecting the above snippet, the reader can notice that the functional properties *model, warrantyYears, colour, cost, operatingSystem* may have values the user is not aware of. If the user does not specify a value for such properties, they are still considered in the matchmaking process as if their value were arbitrary. Thus, for instance, the result of a search for a white-coloured laptop with at least 2-years warranty may include the following results:

```
Laptop#1 M: Sony Vaio W: 2-year Col: white C: 1500$
         OS: ArchLinux 2009.02  SN: 65TG7890
Laptop#2 M: HP TX W: 2-year Col: white C: 1800$
         OS: MacOS SN: 88TY8906
Laptop#3 M: Toshiba W: 3-year Col: white C: 1100$
         OS: ArchLinux 2009.02
Laptop#4 M: Toshiba W: 2-year Col: white C: 1000$
         OS: ArchLinux 2009.02
```

It can be noticed that the results above can be organized in different ways. They can be included in a flat list, as just shown above, or can be visualized in groups and ordered by the additional properties, that they have in their ontology representation, and that have not be used in the search. This second option leads to show the above results as follows:

```
Group#1 (Cost, Model, OS, SN)
-----------------------------
Laptop#1 M: Sony Vaio W: 2-year Col: white C: 1500$
         OS: ArchLinux 2009.02  SN: 65TG7890
Laptop#2 M: HP TX W: 2-year Col: white C: 1800$
         OS: MacOS SN: 88TY8906

Group#2 by (Cost, Model, OS)
-----------------------------
Laptop#3 M: Toshiba W: 3-year Col: white C: 1100$
         OS: ArchLinux 2009.02
Laptop#4 M: Toshiba W: 2-year Col: white C: 1000$
         OS: ArchLinux 2009.02
```

The grouping criteria can be numerous and allow the users to learn the values of the unspecified properties. The grouping used here is by the number and type of additional properties, but other criteria can be adopted. The user may have learnt by simply expanding the first group of results that he was interested in OS Linux laptops only, without further exploring the remaining groups. Thus, he could refine the search by repeating the same query by adding the OS value.

The same structure may characterize several ontologies actually available on Web libraries, having different URIs, such as *http://shopping.yahoo.com/computer.owl* and *http://shopping.ebay.com/mobiledevices.owl*. The user may choose to run the same query on different ontologies and to check the results by different providers. This means that in the visualization of results the provenance of laptops may also be highlighted, as shown in the following snippet, which is output by the distributed version of our matchmaking system.

```
Provider#1 at http://shopping.yahoo.com/computer.owl
---------------------------
Laptop#1 M: Sony Vaio W: 2-year Col: white C: 1500$
         OS: ArchLinux 2009.02  SN: 65TG7890

Provider#2 at http://shopping.ebay.com/mobiledevices.owl
---------------------------
Laptop#2 M: HP TX W: 2-year Col: white C: 1800$
            OS: MacOS SN: 88TY8906
Laptop#3 M: Toshiba W: 3-year Col: white C: 1100$
            OS: ArchLinux 2009.02
Laptop#4 M: Toshiba W: 2-year Col: white C: 1000$
            OS: ArchLinux 2009.02
```

## 2.2 Architecture of the Recommender System

Before delving into the details of the matchmaking process, we outline the main modules of the recommender system, which embeds the matchmaker. We first present a centralized version of our system, in which the matchmaking capabilities are entrusted to a single machine, and then show the distributed architecture, in which the matchmaking function is seamlessly distributed among the peers.

We recall that an electronic marketplace consists of a common vending space, in which a set of providers $P_i$ publish *advertisements* of their own resources, and a set of potential clients $C_j$ issue *queries* demanding a particular good, item or service, and expect *recommendations*. Thus, the common vending space, as shown in Figure 1, is the core of the marketplace in which the matchmaking activity takes place and the matchmaking algorithms are applied in order to output the best recommendations.

While adopting the general architecture above, we have devised a set of modules that collaborate and cooperate to realize the marketplace. Figure 2 shows how the architecture of the marketplace is detailed in our system. We start by highlighting the main modules of the marketplace, which are the following:

- **Ontology Base Manager**: it stores the ontologies used by the Provider to publish its own resources and by the Client interface to formulate queries; in order to guarantee safe operations on such OntologyBMS, an Administrator is responsible of insertions of new ontologies and deletions of existing ones;

- **Resource Base Manager**: it contains the instances of the ontologies stored in the OntologyBMS, that represents the advertisements published by the Provider; the Matchmaker module uses such advertisements to match the user queries;

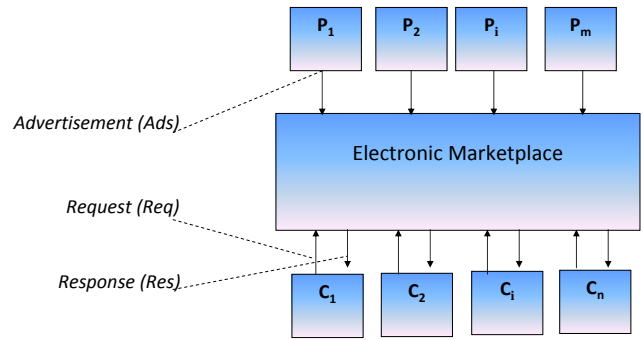

Figure 1: General Architecture of a MarketPlace

- **User Profile Handler**: it is responsible of creating a user profile, by analyzing the past user queries and the user behavior; thus, the queries used in the past can be stored in the user profile base, with a validity field, and can be periodically issued until validity expires; it is responsible of handling a profiler ontology that describes the user preferences;

- **Event Base Handler**: this module allows to capture possible events on the marketplace, such as the publication of new resources that satisfies the user needs, or just a new resource that was not available before;

- **Rule Manager and Rule Engine**: this module and engine are responsible of handling rules in a logical formalism, in order to classify the users into categories and construct suitable user profiles.

- **Matchmaker Module**: it is the core of the system, and is responsible for the matching of the incoming user queries with the advertisements provided by the vendors; it also interacts with the Rule Manager to transmit past queries and build the user profile. It can also capture events raised by the changes on ontologies in order to satisfy the user needs. The matchmaking algorithm that is behind this module will be described in detail in the next section, whereas in the following we focus on the matchmaking strategies that are supported by our system.

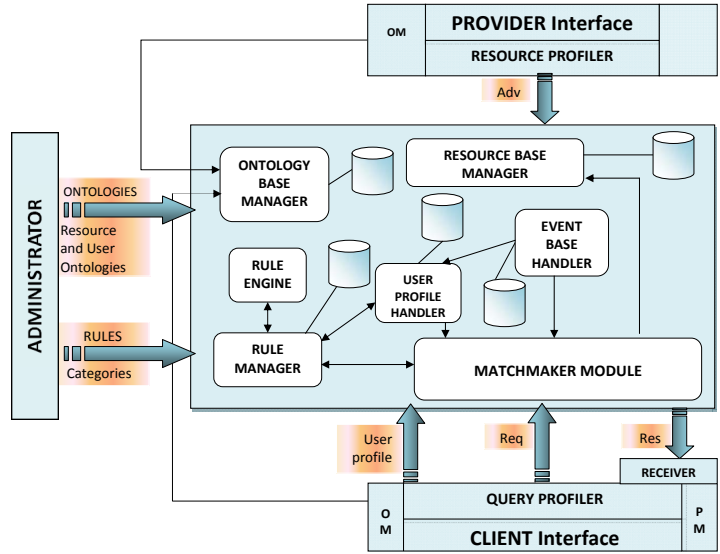

Figure 2: Architecture of a Centralized Recommender System

## 2.3 Matchmaking Options

We have devised two main modalities that drive the matchmaking task, that we have called PULL and PUSH, respectively. The PULL modality is adopted whenever the user explicitly issues a query, by formulating a request for a resource or a set of resources. The client prepares this query through the Query Profiler interface, and receives a response through the receiver. The issued query may be saved in the system for later use, i.e. when new resources are added. This is done by means of the Profile Manager. The queries saved can be used in the second matchmaking option, in which the PUSH modality is applied. During the application of such modality, the user just needs to login and a set of results will be prompted, based on past queries and user profile. In such a case, the saved queries may have been fired by possible changes in the ontology. Similarly, by leveraging a profiler ontology, in which the user preferences have been registered, possible recommendations are prompted to the user. In more details, the user profile consists of two components: a set of queries the user saves and an instance of a 'profile' concept belonging to the profiler ontology. Such an ontology is coupled with a set of inference rules which are handled by the Rule Engine module at the client side and are fired by the user login event. By running these rules, i.e. checking some conditions on the profile elements, the system infers what categories the user belongs to; in such a way, we can recommend resources that have been created and formerly classified according to these categories. Obviously, the inferred categories change if the user changes his/her profile. Nevertheless, the categories may change dynamically if some conditions have been evolving (e.g. a condition on the age of the user).

## 2.4 Clients and providers at a glance

While being compliant to the general architecture of the marketplace, depicted in Figure 1, the providers and clients themselves are further decomposed into modules (for simplicity, only one provider and one client are reported in Figure 2).

Each client consists of a *Query Profiler* that is the user interface to formulate the query using the ontology constructs. Moreover, it has an *Ontology Manager* that retrieves the corresponding ontologies from the OntologyBMS and uses them to prepare the Query Profiler. It also includes a *Profile Manager* that allows the user to explicitly modify its profile, or implicitly applies changes based on the actual preferences of the user on the latest recommendations. Finally, it has a *Recommendation Receiver* that receives the recommendations issued by the Matchmaker and displays them to the user.

Conversely, each provider consists of a *Resource Profiler* that is the user interface to actually fill the resource features as they are imposed by the corresponding ontology. To this purpose, it interacts with the OntologyBMS by means of an *Ontology Manager* that retrieves the corresponding ontologies and uses them to prepare the Resource Profiler.

Besides the providers and clients, a superuser called Administrator is responsible of updating the ontologies in the OntologyBMS and handling the rules in the Rule Engine.

## 2.5 Distributing the matchmaking task

Figure 3 shows the distributed version of our system. In our description, we focus on the intrinsic properties of this architecture and highlight the differences with respect to the centralized version. In Section 5, we will show both architectures at work, by properly tweaking their matchmaking effectiveness and performance.

Figure 3 shows a simplified architecture in which the internal repositories of the event base, resource base, user profile base and rule base have been omitted. We discuss the role and the location of these various repositories at the end of this paragraph. The main advantage of this architecture is the fact that the matchmaking service (and the consequent load) is distributed among the peers. The matchmaker module in such a scenario is a thin module which is easy to deploy on any Ontol-ogyBMS. It is implemented as a Web service, using the SOAP protocol. What is needed on each knowledge base is a WSDL description of the web service, after the provided ontologies have been registered within a common OntologyRegistry. The OntologyRegistry module receives requests for adding and deleting ontologies by the providers and stores values of these ontologies in a dedicated repository. These values consists of a set of keywords, extracted from the ontologies, along with the URI of the corresponding ontologies. The keywords are useful to facilitate the search of the ontologies the user is interested in, in order to formulate his demands. This choice is done through the client interface, through which the user can actually visualize the ontologies present in the OntologyRegistry, and possibly filter them according to the preferred keywords. Once the user has completed his choice, he can formulate his query based on the selected ontology. Later, he can issue this query on this ontology located on the corresponding provider, or, alternatively, on a set of similar ontologies, sharing the same keywords, and located on different providers. In such a case, the user can formulate a single query detailed with respect to a target ontology and then forward it to different providers which share the same terminological structure (TBox) with respect to the target ontology. For the time being, we have restricted to the scenarios in which the providers publish ontologies sharing the same TBox. The latter case turns to be non trivial and poses several challenges. We also observe that more generic scenarios in which the providers publish ontologies with similar, but not identical TBox require to solve the alignment problem between them [5], and is far beyond the scope of our work.

We conclude the discussion about the distributed architecture, by discussing the locations of the event base, resource base, user profile base and rule base in the new architecture. The event base handles the events raised by the changes on the ontologies stored in the OntologyBMS. Thus, this component must be jointly located with each OntologyBMS, and should also capture the user logging events on the client interface in order to enact the PUSH mode. The resource base should also reside on each OntologyBMS, in order to collect the resources of the ontologies. Conversely, the user profile base component is stored on the client side, and is responsible for the management of user profiles that are built according to the profiler ontology, that is also saved in the OntologyBMS. The profiler ontology has here the same role, as in the centralized architecture, i.e. to classify the users interested in its resources. Based on the profiler ontologies, the rule base is capable of running rules that decide the categories to which a user belongs. Since the user profile base also allows to save the queries issued by the user, these can also be exploited in the classification of users by rule inference. As a consequence, the rule base component is also within the OntologyBMS.

To conclude the discussion about the differences between the centralized and distributed architectures of our system, we can observe that while in the central-ized case, the matchmaker and the various repositories are centralized, in the distributed case, they are located on each peer, thus sharing their tasks among several machines. Since the computation is done by several peers at a time, this computation can actually take place in parallel, thus bringing to an improvement of the performance, as we prove in the experimental section.

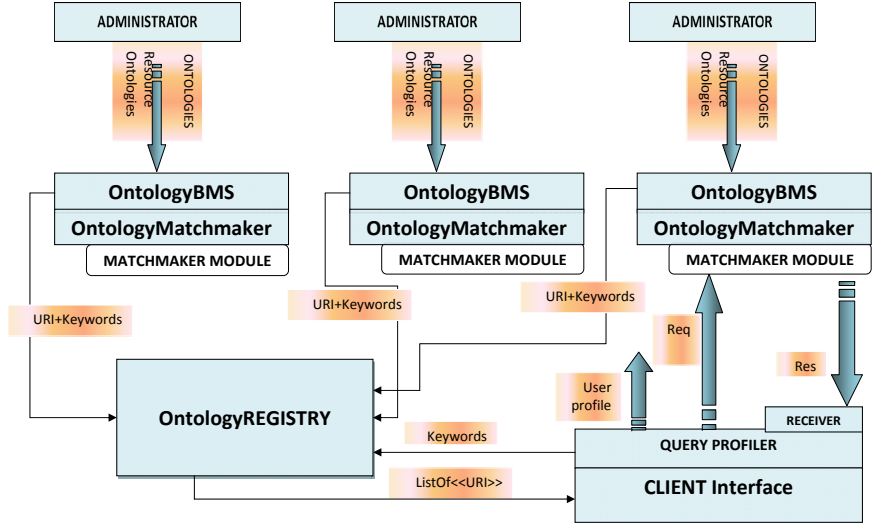

Figure 3: Architecture of a Distributed Recommender System

## 2.6 Comparison with previous architectures

We would like to highlight in this section the differences between our architectures and previous matchmaking prototypes, such as [13, 9]. The latter ones are both agent-based matchmaking systems, that were designed before OWL officially became a standard. [9] represents the first matchmaking prototype that uses a DAML-S ontology and a Description Logics reasoner to compare ontology-based service descriptions. The usage of a DL reasoner implies that the query and advertisement must actually be identical and are subsumed in the specific case by the concept `ServiceProfile`, which is defined in a shared ontology. Although [9] was based on Semantic Web technology, the deployment has been quite different in our case, as we do not assume a mediated ontology in the centralized case, but any ontology base can be plugged in. InfoSleuth [13] is also based on broker agents that a share common vocabulary based on a single domain-specific ontology. Ontologies are not expressed in the standard ontology language, but in a logical deductive language, called LDL++. To the best of our knowledge, none of the previous matchmaking systems exploited could support a distributed matchmaking service.

## 3 The Matchmaking Process

To implement the matchmaking module in both architectures, we have adapted the matchmaking algorithm in [4], with some enhancements. We first describe the matchmaking procedure in [4], which consists of two separate steps, *RankPartial* and *RankPotential*, both targeted to an ontology language with limited expressive power. In these steps, demands and supplies are indeed provided as input logical formulas. These formulas are preprocessed and partitioned into *concept names*, and *constraints*. *RankPartial* applies an initial hard-pruning to the demands that may potentially be satisfied by verifying that they are not disjoint with respect to the supplies. In doing so, *RankPartial* checks the disjointness of the concepts, and the disjointness of the constraints. Finally, the procedure yields as output a ranking value $n_{par}$ for each pair given by a resource in the supplies and the demand. *RankPartial* provides a pre-processing step before applying *RankPotential* that actually computes the matchmaking between demands and supplies, by only considering those that are not disjoint. Similarly to *RankPartial*, *RankPotential* inspects the concepts and better ranks those concepts (constraints, resp.) in the supply that are strictly contained or coincide with the demand. *RankPotential* also outputs a ranking value $n_{pot}$ for each pair given by a resource in the supplies and the demand, by taking into account the number of overlapping concepts between demand and supply.

We thought that the above algorithm as in [4] was appropriate for a matchmaking analysis between ontology-based resources. Indeed, other algorithms, such as the one in [9], lacked the concrete datatypes and only considered demands and supplies in a common shared representation.

However, we needed to extend the algorithm in [4] to cover a broader number of OWL constructs, and to guarantee that the matchmaking between a demand and a large number of supplies is meaningful and the results are optimally ranked. Other changes were actually performed in order to apply useful optimization steps. We summarize the applied changes in the following list and refer the reader to the complete pseudocode in Fig. 4:

- **extension for OWL classes and properties:** we have customized the matchmaking algorithm to work with the ontology-language standard, i.e. OWL, a prerequisite for the development of the Semantic Web. OWL is based on a very expressive DL called *SHOIN(D)* [7], that includes classes, properties, class constructors and hierarchies of classes and properties, which we are able to cover in our framework;

- **lower number of comparisons:** Steps **2** and **3** of the Algorithm are devoted to compute the values of $n_{par}$, $n_{pot}$ and $n_{add}$ for the resources in the supply set $S$, and their values for the constraints. This computation requires a number of comparisons equal to $|D| * |S|$, where $D$ is the set of demands; however, in many cases, the number of comparisons can be reduced, as we discuss in Section 3.1;

- **knowledge elicitation:** the ranking values $n_{par}$ and $n_{pot}$ were not sufficient in all the scenarios, in which the user has to be guided in the list of matched results by additional properties of resources; for instance, we have enhanced the algorithm by adding a third ranking value, called $n_{add}$, representing the number of additional concepts a resource has in the supply, that have not been specified in the demand; this allows to enhance the knowledge elicitation and better compute a *unified ranking value* as done in Step **4**.

## 3.1 Comparing demand and supply

Steps **2** and **3** of the Algorithm perform a double inspection of the concepts in the demand and of the concepts in the supply. We realized that the number of comparisons could be further reduced by applying some pre-processing. Moreover, the double inspection of concepts in Step **2** and of constraints in Step **3** can actually be done in parallel. Indeed, the *disjoint* relationship between each concept in the supply $S$ and the concepts in the demand $D$ corresponds to a total number of comparisons, given by $|S| * |D|$. The *not among* relationship actually does the same number of comparisons, albeit in a reverse order. In order to avoid making a large number of comparisons, some optimizations can take place. Figure 5 shows a supply consisting of two concepts $\{D, E\}$ and a demand consisting of three concepts $\{A, B, C\}$. First of all, the disjointness condition between each concept in $\{D, E\}$ and each concept in $\{A, B, C\}$ implies that each pair of concepts must be disjoint. If concept $D$ and concept $A$ are disjoint, then the opposite holds. This leads to observe that the *disjoint* relationship is indeed symmetric. The *not among* relationship has to be applied to two sets of concepts (e.g. $\{A, B, C\}$ and $\{D, E\}$ in the example), and aims at checking that the concepts in the first set do not belong to the second set. For instance, in Figure 5, the concepts in $\{A, B, C\}$ are not among the concepts $\{D, E\}$. Whereas the *disjoint* condition can be computed by intersection (i.e. by checking the emptiness of the intersection), the *not among* condition

**Algorithm SemanticMatchmakingAndRanking.**
*Input*: ontology concepts in demand $D$ and supply $S$.
*Output*: supply $S$ with ranking function $rank$.
**1**. Let $n_{par}$ be the number of concepts in $S$ disjoint from the concepts in $D$;
       $n_{pot}$ be the number of concepts in $S$ not among the concepts in $D$;
       $n_{add}$ the number of additional concepts in $S$ w.r.t. $D$
       $S_{names}$ the concept names in supply $S$;
       $D_{names}$ the concept names in demand $D$;
  For each supply $s_i \in S_{names}$
     $n_{par} := 0, n_{pot} := 0, n_{add} := 0$
**2**. For each supply $s_i \in S_{names}$
       For each concept $c_i \in s_i$, having weight $c_{i_w}$:
       if $c_i$ is disjoint from the concepts in $D_{names}$
         $n_{par} := n_{par} + c_{i_w}$
       Add to $n_{pot}$ the number of concepts in $D$ that are not among those in $s_i$
       $n_{pot} := n_{pot} + |D_{names} - S_{names}|$
       For each concept $c_j \in s_i$
         $n_{add} := n_{add} + numberAdditionalConcepts(c_j, D_{names})$
**3**. For each constraint $n_{c_i}$ of concept $c_i \in S_{names}$, having weight $c_{i_w}$
       if $n_{c_i}$ is in conflict with a constraint in $D_{names}$
         $n_{par} := n_{par} + c_{i_w}$
  For each constraint $n_{d_i}$ of concept $d_i \in D_{names}$;
       if $n_{d_i}$ is not implied by any constraints in $C_{names}$
         $n_{pot} := n_{pot} + 1$
**4**. Let $rank_{par} := 0$ be the ranking based on $n_{par}$;
       $rank_{pot} := 0$ be the ranking based on $n_{pot}$;
       $rank_{add} := 0$ be the ranking based on $n_{add}$;
       $max(n_{par})$ the maximum value of $n_{par}$ for supplies $\in S_{names}$;
       $max(n_{pot})$ the maximum value of $n_{pot}$ for supplies $\in S_{names}$;
       $max(n_{add})$ the maximum value of $n_{add}$ for supplies $\in S_{names}$;
  For each supply $s_i \in S_{names}$
      if $max(n_{par}) \neq 0$
         $rank_{par} :=$ normalize $n_{par}$ wrt. $max(n_{par})$
      if $max(n_{pot}) \neq 0$
         $rank_{pot} :=$ normalize $n_{pot}$ wrt. $max(n_{pot})$
      if $max(n_{add}) \neq 0$
         $rank_{add} :=$ normalize $n_{add}$ wrt. $max(n_{add})$
      Output $rank = StdMean(rank_{par}, rank_{pot}, rank_{add})$

Figure 4: **Matchmaking Algorithm with Knowledge Elicitation**

can be computed by difference of the two sets and it holds true if the result of the difference coincides with the first set.

We have introduced a pre-processing phase in the algorithm that takes into account the commonalities in the algorithm steps and avoids redoing the same comparisons. The total number of comparisons is thus reduced by an half (i.e. is $|S| * |D|$ at most). In particular, the *disjoint* condition can be computed by difference as well, by exploiting the equivalence $A \cap D = A - (A - D)$, and once computed the disjoint condition, these can be also be exploited in the computation of the *not among* condition, as follows. In the example, the concepts in $\{A, B, C\}$ are not among the concepts $\{D, E\}$ iff each concept in $\{D, E\}$ is disjoint from each concept in $\{A, B, C\}$. The figure shows that the computation of the not among relationship just exploits the comparisons already done for the disjoint relationship.

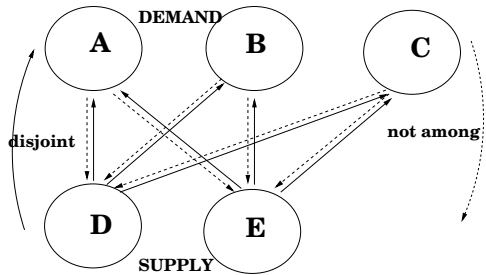

Figure 5: Comparing Demand and Supply

## 4 Ontology-driven Result Visualization

Besides enabling the matchmaking function, a recommender system should also be able to visualize the results to the users in an understandable fashion, and rank them according to some criteria.

Our system allows the user to specify a confidence value in their demands, for instance on properties. In our running example, the request for a white-coloured laptop, can be formulated by specifying a confidence value from 1 to 10, which actually tells how compulsory has to be a predicate. In such a case, a confidence value equal to 3 means that the user is interested to this condition for $30\%$, and he can give up this restriction for $70\%$.

If no confidence value is specified, a default value equal to 10 is used, meaning that a condition is mandatory for the user. Nevertheless, the matchmaking algorithm is able to compute a ranking value for each resource, based on the number of additional properties, $n_{add}$, of resources and the usage of partial and potential ranking values, $n_{par}$ and $n_{pot}$, as explained in the previous section.

Besides ranking, the system lets exploit two possible strategies for the visualization of results. The first strategy, which we call *Naive*, allows to visualize the results in a flat list, by putting on the top of the list the results that more closely satisfy the user demand (included the confidence value, if this has been specified). This flat list assumes the ranking values as computed by the matchmaking algorithm.

By expanding the items of the list, the other properties of a resource can be visualized, i.e. those that have not been specified in the demand, but are part of the ontology representation of that resource in the knowledge base. However, we realized that this strategy poorly helps the user in the process of result visualization, as he/she has to browse the entire list to learn the possible additional properties of a resource, and the possible values of this property.

Given the limitations of the naive strategy, we devised a second strategy, called *Grouping*, that guides the user during the inspection of the results. Here, we allow the organization of the output results into 'groups'. Each group is built on the number and type of properties not specified in the demand in order to improve the expressiveness of the user queries, as shown in the use case in Section 2.1. These properties are the same used by the algorithm in the $n_{add}$ parameter, thus have been already considered in the computation of the ranking value. In this strategy, they are used for visualization purposes. By using the *Grouping* strategy, the results are split into groups, each group having one, two and, accordingly, an arbitrary number of unspecified properties. The user can now expand the group(s) that he likes and possibly use them to refine his query. For instance, by using our running example, a user may have issued a first demand on a black wide-screen PDA. He does not know what would be the possible values of other features, such as operating system, CPU size, RAM etc., that are for the time being left blank in formulating the demand. By using the *Naive* strategy, the system would output the results in a flat list, with a specified ranking order. However, the results may be too numerous to be inspected by the user. By repeating the same query with the *Grouping* Strategy, the results would be organized in a convenient way. This result visualization allows to group the output resources according to their additional properties, and lets the user browse within these groups in order to find the best resources at all. In the example discussed in Section 2.1, the first group shows the supplies that only have the properties that have been specified in the demand 'black wide-screen PDA', thus with zero additional properties, whereas the second group shows the supplies having one additional property (for instance, OS or CPU size), the third group shows the supplies with two further additional properties and so on. The user may thus decide to expand the groups in order to learn the values for the additional properties of the requested object. As an example, the user learns that the available PDAs may have a Windows or Linux OS.

As part of the user study, we have performed a user evaluation of these two strategies and their usefulness in the matchmaking process. This will be illustrated in the next section.

## 5 Experimental Study

We have conducted an experimental study on both the centralized version and the distributed version of our prototype. The experiments aimed at investigating the suitability of both architectures with real-life ontologies and matchmaking cases. Such investigation was twofold, as it was expected to measure the performance of the matchmaking prototype, and the effectiveness of the matched results. We thus first arranged a user study, by letting a group of users working on the prototype and asking them to issue their demands and filling a questionnaire, based on the results of the matchmaking process. Finally, we measured the execution times of the matchmaking algorithm in the centralized case, and that of the matchmaking algorithm in the distributed case.

We have implemented both architectures in Java 6.0, by using PostgreSQL 8.2 as the back-end for the ontology management (OntologyBMS) and MySQL as the back-end for the ontologyRegistry, Jena2 as the knowledge base, the rule base and the user profile base (including user's queries). Jena2 Semantic Framework stores ontologies as RDF triples by providing a layer on top of a relational back-end. Querying the ontologies is done by means of a SPARQL engine. Finally, Apache Tomcat 6 has been used as Server Container while Axis2 as a module to enable the SOAP/XML interaction with the OntologyRegistry, and to wrap requests and responses.

### 5.1 Experimental Setting

Throughout the experiments, we have used a set of ontologies, that represents real purchases of goods or services, such as $Computer.owl$, and $Book.owl$ for goods, and $Doc - EGov.owl$ for public administration services. Moreover, we have put our system at work on a standard ontology, $Wine.owl$ [1], as provided by the World Wide Web Consortium.

Table 1 shows for each ontology its main characteristics in terms of number of concepts and number of properties used in formulating the demands.

---

[1]The $Wine.owl$ ontology is publicly available at the following url: *http://www.w3.org/TR/2004/REC-owl-guide-20040210/wine.rdf*.

| Ontology | # concepts | # objectProperty | # datatypeProperty |
|---|---|---|---|
| Computer.owl | 13 | 6 | 0 |
| Books.owl | 9 | 2 | 13 |
| Doc-EGov.owl | 22 | 10 | 67 |
| Wine.owl | 80 | 12 | 9 |

Table 1: Ontologies used in the experiments and their main features.

The experiments on the centralized architecture have been executed on a Windows XP Laptop with 2.00 GHz, and 1GB memory. The experiments on the distributed architecture have been executed on Windows XP Pro Laptops with memory ranging from 512MB to 4GB. The ontologies have been replicated on each machine, and the results have been collected by one of the machines, acting as a client and as an OntologyBMS at the same time.

### 5.2 A User Study for Assessing the Effectiveness of the Matchmaking Algorithm

We have asked a group of 16 users to utilize our system, and fill in a questionnaire with a set of questions, in order to measure the user satisfaction in the various matchmaking cases. While choosing the demands that the users used in the study, we opted for real-life demands of goods and services. Our user study aimed at actually demonstrating the usefulness of the PUSH and PULL modality of our system, and the adequacy of the grouping strategies.

We conducted our study as follows. We employed two of the ontologies shown in Table 1, precisely $Doc - EGov.owl$ and $Computer.owl$. The reason why we chose these two ontologies is due to the fact that our users were chosen among the students enrolled in the Computer Science Program at University of Basilicata, thus representing a sample of users more acquainted with computers and less acquainted with public administration services. We will later observe that this choice actually confirms that the familiarity of such users with computer goods did not bias the user study study results and confirmed the usefulness of the two matchmaking strategies.

For each ontology, the users were asked to use the system by issuing a query. We employed two queries for $Doc - EGov.owl$ ontology and three queries for $Computer.owl$ ontology. The first two queries on $Doc - EGov.owl$ ontology were on tax payments on behalf of the administrative staff of the Italian Internal Revenue Service. The first two queries on $Computer.owl$ ontology were on computer purchases with different properties, whereas the latter query was a query on the number of laptops, satisfying a particular condition, for instance done for sta-tistical purposes. A description of each query is shown in Table 2.

| Ontology | Query | Query Description |
|---|---|---|
| Computer.owl | QC1 | The user is interested to buy and retrieve the laptops with price less than 1000 euros and with 2 years-warranty or more |
| Computer.owl | QC2 | The same as above, but with a confidence value for the second predicate |
| Computer.owl | QC3 | The user wants to build a statistic on the number and type of grey-coloured laptops, with price equal or greater than 1000 euros |
| Doc-EGov.owl | QD1 | An IRS employee retrieves the data about citizens with Debit Payment greater than 25 euros. |
| Doc-EGov.owl | QD2 | A bank officer retrieves the tax payments due on 02/27/2009, with amount greater than 150 euros |

Table 2: Queries of the user study.

Based on these queries and ontologies, we asked the user to fill a questionnaire, by answering a set of questions, divided into boolean questions and questions with a numeric range, representing the level of satisfaction from 1 (low) to 3 (high). The results of the user study are reported as the following measures:

- the average value of user satisfaction $\%US$, indicating the average of positive answers for boolean queries and the average of high-rated answer for range queries, respectively;

- the standard deviation of user satisfaction $\%SD$, indicating the spread out of values.

The first step in our user study was to actually probe the effectiveness of the PUSH mode. We recall that in the PUSH mode, the user is asynchronously provided with a set of resources, that are his/her preferred ones. The user profile is employed in this mode, along with the past queries issued by the user. We have asked the users to fill in their user profile and to enter a set of queries. Later on, the next time the users logged in, they were prompted with their preferred sets of resources, that were newly added to the knowledge bases. We asked the users the following questions, as shown in Table 3, in which we also show the average $\%US$ of satisfaction and the standard deviation $\%SD$.

The experiment was carried out on $Doc - EGov.owl$, which contains services of the public administration. As shown in Table 3, the conclusion of the user study on the PUSH mode was that the user satisfaction is quite high, with a fairly low spread out of values.

| Question | %US | %SD |
|---|---|---|
| Do you deem useful the PUSH mode? | 74,4 | 19,2 |
| Are you satisfied with the (PUSHed) resources? | 76,9 | 24,1 |
| Are the (PUSHed) resources compliant with your profile? | 92,3 | 26,6 |

Table 3: Summary of Results of the User Study (PUSH mode) on $Doc - EGov.owl$.

| Question | Acr. | Strat1 | Strat2 |
|---|---|---|---|
| Do you deem useful the Provided Ranking? | UPR | A | A |
| Are the most significant results in Top-3 position? | STK | A | A |
| Do you deem useful the Ranking wrt. Additional Properties? | RAP | NA | A |
| Do you deem useful the Grouping wrt. Additional Properties? | GAP | NA | A |

Table 4: Most Significant Questions of the User Study (PULL mode), their acronym and their applicability to the two Strategies.

The user study then proceeded to analyze the PULL mode. In such a case, we asked the users a number of questions on their satisfaction in the various cases, when they explicitly type a set of demands. In particular, in order to collect the users opinion on the *Naive* and *Grouping* strategies, we asked them to repeat the same query under two possible anonymous strategies (Strat1 and Strat2, standing for Naive and Grouping, respectively) and write down their rating in both cases. We preferred to not tell the user the name and details of the strategies to avoid biasing the study. The user study for the PULL mode consisted of a series of questions in a questionnaire. For the sake of presentation, we have summarized the questions into a number of four most significant questions we posed to the users, and which we report in Table 4. For each question, we present its description, an acronym that we will use in the remainder of this paragraph, and its applicability to the two strategies. Whereas the first two questions focus on the ranking, thus are applicable to both strategies, the latter two questions concerns the usefulness of additional properties ($n_{add}$), thus are only appropriate for the second strategy.

Table 5 shows the results of the user satisfaction and the standard deviation for the questions above, these results being grouped by the corresponding strategy. It can be noticed that in all queries, as shown by the answer to question $UPR$, the average of user satisfaction $\%US$ is greater for the second strategy than for the first. This shows that the ranking provided by the second strategy, by highlighting the number of additional unspecified properties, is deemed more useful from the users' point of view that a simple ranking without grouping. This is also confirmed by the high percentage of $\%US$ for the answers to questions $RAP$ and $GAP$. The only exception is represented by query $QC3$ on $Computer.owl$ (cfr. Table 2), which was aimed at building a statistics. In such a context, the question $GAP$ shows that, whereas the ranking is deemed useful (cfr. $RAP$ for $QC3$), the grouping is fairly not helpful, and confirmed our assumption that the grouping really helps the user when he is interested to buy a good or service (i.e. for all queries other than $QC3$), and is not familiar with its various unspecified properties. If we look at the results for question $STK$, we can observe that the first strategy is deemed better than the second strategy as a means to show the Top-3 positions of the results list for queries $QC1$ and $QC2$. The results are reverted for the following query $QC3$, $QD1$ and $QD2$, where the second strategy turns out to be better than the first one. The conclusion is that this was somehow expected, since the users, being computer scientists, feel more confident with the first two queries, that are purchases of computers, whereas they are not much confident with the remaining ones, and, in the latter cases, Strat2 with grouping was deemed more helpful to identify the best Top-3 results.

| Query | Question | %US(Strat1) | %US(Strat2) | %SD(Strat1) | %SD(Strat2) |
|---|---|---|---|---|---|
| QC1 | UPR | 66,7 | 69,2 | 26,1 | 30,6 |
| QC1 | STK | 82,1 | 72,2 | 24,9 | 29,9 |
| QC1 | RAP | - | 88,9 | - | 22,2 |
| QC1 | GAP | - | 83,3 | - | 37,3 |
| QC2 | UPR | 61,5 | 69,2 | 22,1 | 20,5 |
| QC2 | STK | 69,2 | 66,7 | 24,3 | 26,1 |
| QC2 | RAP | - | 87,5 | - | 23,2 |
| QC2 | GAP | - | 83,3 | - | 37,3 |
| QC3 | UPR | 69,4 | 77,8 | 28,7 | 28,3 |
| QC3 | STK | 72,2 | 86,1 | 29,9 | 25,3 |
| QC3 | RAP | - | 87,5 | - | 16,1 |
| QC3 | GAP | - | 27,8 | - | 12,4 |
| QD1 | UPR | 52,1 | 68,8 | 20,3 | 24,9 |
| QD1 | STK | 64,6 | 68,8 | 18,5 | 27,6 |
| QD1 | RAP | - | 76,7 | - | 21,3 |
| QD1 | GAP | - | 81,3 | - | 39,0 |
| QD2 | UPR | 53,3 | 64,4 | 20,4 | 25,7 |
| QD2 | STK | 62,2 | 64,4 | 23,9 | 25,7 |
| QD2 | RAP | - | 70,0 | - | 31,4 |
| QD2 | GAP | - | 66,7 | - | 47,1 |

Table 5: Summary of Results of the User Study (PULL mode).

## 5.3 Performance Analysis

We have analyzed the performance of our system, by utilizing the centralized version and the distributed version. Figure 6 reports the execution times of a set of demands that were run on each ontology in the centralized architecture. For the sake of comparison, we have filled each knowledge base with the same number of instances and run various kinds of queries, by incrementally varying the number of concepts and properties used in the demands. The queries used in the tests are reported on the $X$ axis of the plot in Figure 6, along with the number of properties specified in each query. For instance, queries *Q1-1* and *Q1-2* indicates the same query *Q1* with one property specified in the first case, and two properties in the second case. We have used an incremental number of properties in the various queries, whereas each query has been built around a single concept of the ontology.

Each result shown in Figure 6 has been averaged on three hot runs of the matchmaking engine. We can notice that for all the ontologies, the execution times are reasonably low. In particular, for $Book.owl$, $Computer.owl$ and $Wine.owl$, the execution times for matchmaking are below 1.4, 1.2 and 2.5 ms, and scales linearly with the number of properties used in the queries. Indeed, the total number of properties for such ontologies, as shown in Table 1 is not high, thus the inspection of additional properties does not impact the times. If, instead, we observe the results for $Doc - EGov.owl$, we can notice that the execution times blow up (till 7 ms). This is due to the fact that this ontology has a greater number of properties. Despite this, the linear scale is preserved for all the queries of this ontology, similarly to the previous ontologies.

Figure 7 shows the results on our distributed architecture. In Figure 7 (LHS), we show the time composition into network latency time and matchmaking time for a set of queries. Besides the number of properties, we also indicate the number of machines employed in the experiment, to name the queries (e.g. query *Q1-2-2* is a query with 2 properties, executed on 2 machines). In Figure 7 (RHS), we show for each experiment the total number of resources matched by the algorithm, that is obtained by summing up the resources matched on each machine. We employed a total of 7 queries in this experiment, where the first three queries are formulated on $Doc - EGov.owl$ ontology, and the remaining ones on $Computer.owl$. We executed two hot runs of each experiment, and averaged the results. If we observe the LHS plot, the execution times of the first three queries on $Doc - EGov.owl$ (in blue) shows a linear scale with respect to the number of properties in the query and the number of machines employed in the distributed setting. The higher is the number of properties and the number of peers, the bigger is the execution time of the matchmaking algorithm. The network latency time is influenced by the number of machines, and by the number of resources (RHS plot), rather than by the number of properties specified in the query. Indeed, the highest latency occurs with queries *Q2-1-2* and *Q3-3-2*, that handle an higher number of resources with respect to *Q1-2-2*. If we look at the last four queries on $Computer.owl$, we can observe a similar trend. The queries *Q4-3-2* and *Q5-1-2* have been executed on two machines, whereas *Q4-3-3* and *Q6-3-3* have been executed on a total of three machines. The first two queries, *Q4-3-2* and *Q5-1-2*, show execution times of the matchmaking algorithm, which only depend on the number of properties, whereas the network latency time is affected by the number of resources that have been shipped around the network, and by the number of machines. The last two queries, *Q4-3-3* and *Q6-3-3*, have both three properties, and have been executed by using a similar number of resources, and the same number of machines. It can be noticed that the matchmaking execution times and latency times are roughly comparable. If we compare the times of query *Q4-3-3* with its previous formulation *Q4-3-2*, we can confirm the previous observation that the matchmaking times are only dependent on the number of properties, and not on the number of resources, while the opposite holds for latency network time.

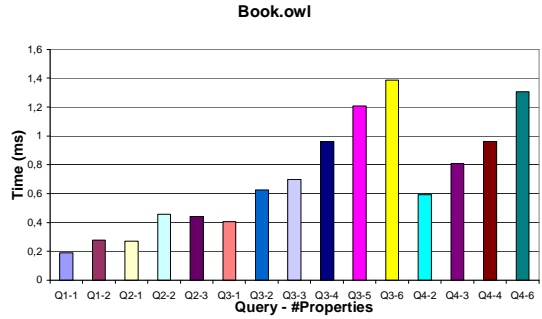

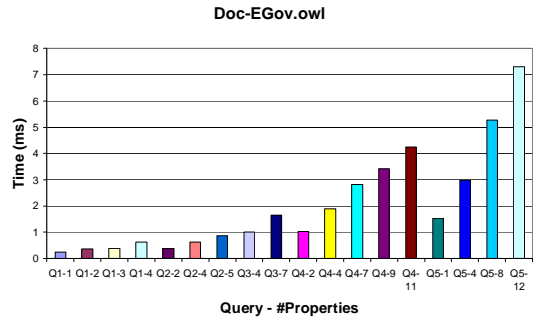

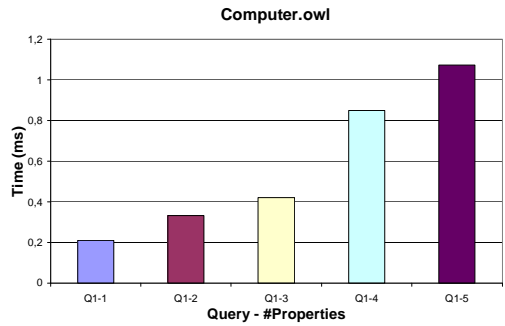

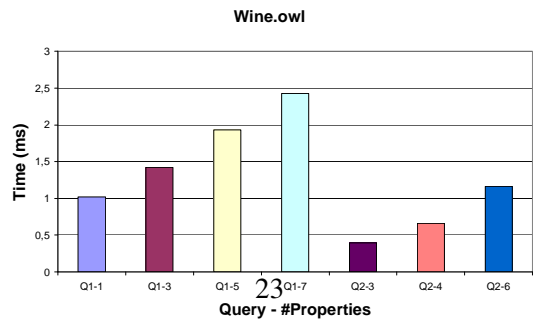

Figure 6: Matchmaking execution times (in ms) in the centralized case.

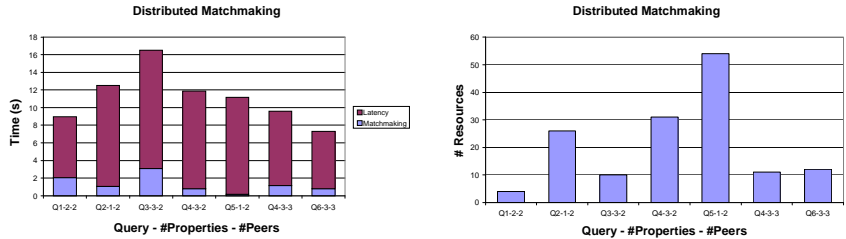

Figure 7: Matchmaking execution times vs. latency network times in the distributed case wrt. #properties,#peers (LHS) and #resources wrt. #properties,#peers (RHS), resp.

In order to improve the total execution time across several machines, we have built an alternative implementation of the matchmaking web service in an asynchronous fashion. While in the synchronous modality, the results coming from each machine are pipelined and gathered on a single machine, in the asynchronous mode, the results are collected in parallel, by exploring multi-thread programming. Thus, in the synchronous mode, the total execution time is obtained by summing up the local times of the machines (i.e. the matchmaking time and latency network time), whereas in the asynchronous mode, the execution is parallelized and the total execution time roughly amounts to the maximum among the local times of the machines. We have thus performed an additional experiment that measures the scalability of the matchmaking service in both modes.

Figure 8 (LHS) shows the total execution time for query $Q4$ in both the synchronous and asyncronous mode by varying the number of machines from 7 down to 3. Figure 8 (RHS) shows the # of resources matched in the various cases. We can notice that the total execution time of the Asynchronous mode ($AS$) is much lower than that of the Synchronous mode ($S$) in all cases. Moreover, while in the Synchronous mode ($S$) the total execution time heavily depends on the number of peers, it is not varying for a number of machines greater than $4$ in the Asynchronous mode ($AS$).

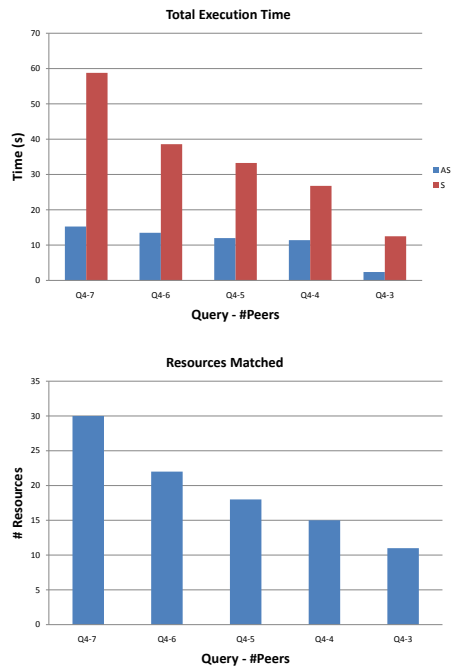

Figure 8: Total execution time in the distributed case wrt. #peers (LHS) and # resources wrt. #peers (RHS).

## 6 Related Work

We have highlighted in the previous sections the differences between our work and the existing matchmaking systems [13, 9, 4]. Although recommender systems are fairly orthogonal to our work, we briefly discuss them in the following. The use of ontologies for user profiling in recommender systems has been studied in [11], which presents an hybrid recommendation approach, both content-based and collaborative. Two recommender systems, Quickstep and Foxtrot, are built for user profiling in the computer science community. In the latter system, ontologies are visualized to allow elicitation of user profile feedback. While the above work focuses on the use of ontologies for user profiling, our focus is on adopting such use all along the matchmaking process.

Yahoo! Vibes [12] is a platform for building recommender systems for large-scale datasets. Contrarily to previous approaches that were customized to a few domains and were not scalable, Vibes is used to power recommendations across a wide range of Yahoo! properties, such as Shopping, Travel, Real Estate etc. It plugs in several recommendation models based on affinity, attribute-similarity and collaborative filtering, and can be extensible to other models. Similarly to our distributed multi-module architecture, it is also based on Web services, to ensure the portability and independence from the language and operating systems. However, Vibes does not rely on ontologies as a powerful means to extract semantics from the items. Instead, it exploits the item-to-item affinity and user historical behavior on Yahoo to build the recommendations. Another limitations is that it can only be deployed for Yahoo items and is not applicable to a wide range of domains.

Amazon [10] and Google News [3] also make use of collaborative filtering, that aims at learning user preferences and making recommendations based on user and community data. Quality of recommendations notwithstanding, they focus on scalability of the recommender system with several million users. Google News also addresses the problem of item churn, i.e. the frequent insertions and deletions of items.

All the above algorithms assume a history of user past clicks that are captured by the collaborative filtering models. Our system does not rely on this assumption, thus being more a content-based (or keyword-based) approach [1] in a loosely fashion. The content we can handle is given by the knowledge and semantics present in ontologies, that are the core of Web 2.0. Thus, the above approaches are orthogonal to ours.

CourseRank [8] focuses on the definition of recommendation workflows, in which recommendations can be customized through a recommend operator and a blend operator is used to unify recommendations provided by different paths. Such operators, together with select, project, join operators allow to build flexible recommendation workflows.

X.QUI.SITE (Yahoo!) [2] is a recommendation platform, built upon a social tagging web site, Yahoo! del.icio.us, through which the users tags the URLs of interest and receives recommendations on hot topics and interesting people. User-centered recommendation is pursued in this system, that aims at presenting the right information to the user. Our system does not assume that the user explicitly tags the information, and we build upon the fact that the users declaratively feed their user profile, while using the latter in the PUSH mode. However, we believe that our system is flexible enough to be extended to social tagging web sites. We also think that the use of ontologies in this context is largely unexplored, and may bring interesting challenges.

# 7 Conclusions and Future Work

We have presented a full-fledged matchmaking system leveraging the use of ontologies to yield and rank the recommendations, and help the users to better reformulate their queries. We have deployed two architectures of our system, a centralized one and a distributed one. We have performed an extensive evaluation, aiming at collecting the user feedback on the various features, and analysing the performance in both architectures.

As future work, we would like to extend our system to utilize community data and collaborative work, for instance to enhance the user profiling (PUSH mode), in the spirit of [11].

Moreover, we would like to investigate the case in which the target ontology that is used by the providers in the distributed architecture, is not unique. Such a scenario would lead to study the complex problems of mapping and alignment of ontologies, that is the subject of a recent book [5].

Moreover, we would like to explore the direction of using our distributed matchmaking services in large-scale loosely coupled architectures, such as in P2P networks.